\documentclass[pre,floatfix,twocolumn]{revtex4}
\usepackage{amsmath}
\usepackage{amsfonts}
\usepackage{mathrsfs}
\usepackage{epsfig}
\usepackage{graphicx}
\usepackage{dcolumn}
\usepackage{hyperref}
\usepackage{xcolor}
\usepackage[hyphenbreaks]{breakurl}
\usepackage{float}
\usepackage{hyperref}
\usepackage{doi}

\begin{document}
\title{Anti-modular nature of partially bipartite networks makes
them {\em infra small-world}}
\author{Aradhana Singh$^{1}$, Md. Izhar Ashraf$^{1,2}$ and
Sitabhra Sinha$^{1,3}$}
\affiliation{$^1$The Institute of Mathematical Sciences, CIT Campus,
Taramani, Chennai 600113, India.\\
$^2$ BS Abdur Rahman Crescent Institute of Science \& Technology, 
Vandalur, Chennai 600048, India.\\
$^3$Homi Bhabha National Institute, Anushaktinagar, Mumbai 400094,
India.
}
\date{\today}
\begin{abstract}
Strong inter-dependence in complex systems can manifest as
partially bipartite networks characterized by interactions
occurring primarily between distinct groups of nodes (identified as
modules).
In this paper, we show that the anti-modular character of such networks,
e.g., those defined by the
 adjacent occurrence of alphabetic characters in corpora
of natural language texts,
can result in striking structural properties
which place them outside the well-known regular/small-world/random network
paradigm. Using an ensemble of model networks whose modularity can be
tuned, we demonstrate that strong module size heterogeneity in anti-modular
random networks imparts them with higher communication 
efficiency and lower clustering than their randomized counterparts, making
them {\em infra small-world}.
Passage to anti-modularity is associated with
characteristic changes in spectral properties of the network, including a
delocalization transition exhibited by the principal eigenvector (PEV)
of the normalized Laplacian. This is accompanied by the emergence of
prominent bimodality in the distribution of PEV components, which can function
as a signature for identifying
anti-modular organization in  
empirical networks.
\end{abstract}

\maketitle

Many complex systems that occur around us can be
described in terms of a network of large number of interacting 
components~\cite{Newman_book,Barabasi_book,structure_dynamics_book}. 
The connection topology characterizing these 
interactions for different systems is one of the most
important factors determining their properties. A dominant paradigm in 
this context
is that of small-world (SW) networks~\cite{Watts1998,Newman2000,Vespignani2018}
spanning a wide range of possible 
topological structures that are distinguished by the global property
of communication 
efficiency (measured by the harmonic mean of all pairwise distances between
the constituent nodes)~\cite{Latora2001} and the local property of clustering
(quantified as the ratio of the numbers of connected node triads to potential 
triads)~\cite{Newman2009}.
Lattices or regular networks having low efficiency, as well as, high 
clustering and Erd\"{o}s-Renyi (ER) random networks, which show high efficiency 
with low clustering, form the two well-understood extreme limits of the 
range of structures encompassed by this paradigm.
Small-world properties have subsequently been reported for a broad
range 
of empirical networks (see, e.g., Refs.~\cite{Albert1999,Newman2001,Wagner2001,
Bullmore2009}).
Indeed, SW networks are far more general
than the context of interpolation between regular and random networks 
in which they were originally
proposed~\cite{Watts1998}. 
In particular, modular networks, characterized by the existence
of subnetworks within which connection density is
significantly higher than that for the entire system,
have been shown to exhibit small-world properties~\cite{Pan09}. 
It is therefore of interest to ask if there are networks 
which fall outside this paradigm, or more aptly, whether 
the class of small-world networks is itself
part of an even more general framework for describing 
connection topologies.

\begin{figure}
\centerline{\includegraphics[width=0.99\columnwidth]{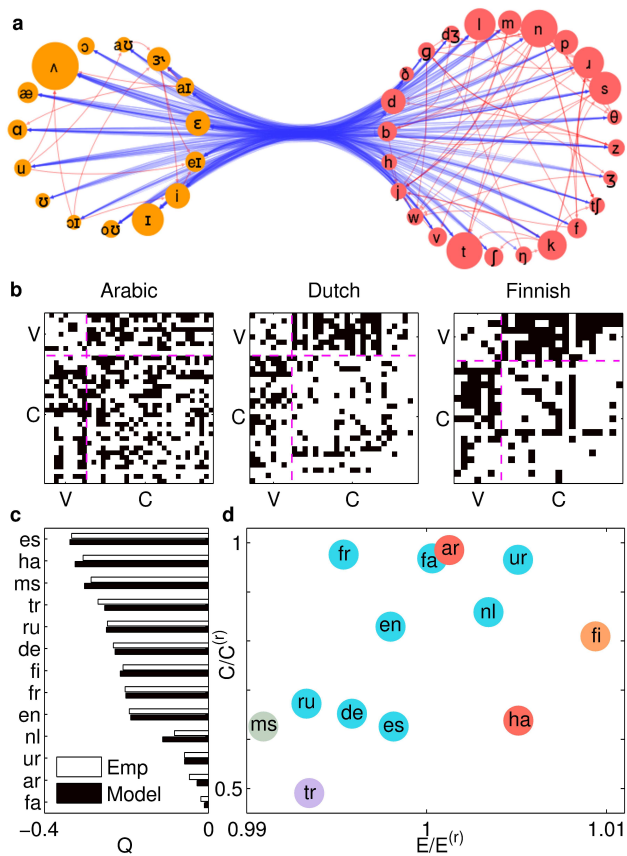}}
\caption{
{\bf Empirical networks with anti-modular character that occur in the context of human languages.}
(a) Spectral partitioning of a phoneme network for English 
reveals an almost
bipartite structure characterized by dense inter-modular and
sparse intra-modular connections 
between one module
comprising only vowels (left) and the other consisting 
exclusively of consonants (right).
Nodes correspond 
to phonemes (indicated by standard IPA
symbols), with directed links representing
statistically significant consecutive occurrence in a word~\cite{note1}. 
Node size represents relative occurrence frequency of
phonemes in the corpus.
(b) Adjacency matrices representing orthographic networks, connecting
letters that occur significantly often at consecutive positions 
in Arabic (left), Dutch (center) and Finnish (right) words.
Note the relatively higher density
of connections between the two modules comprising vowels (V) and
consonants (C), respectively, as compared to
connections within each module.
This suggests an anti-modular organization of the orthographic networks
for a variety of languages that use an alphabetic writing 
system, which is confirmed in (c)
by negative values of the modularity index $Q$~\cite{Newman_spectra}, 
for both the empirical networks (white bars) as well as model networks (see text
for details) having
similar anti-modular structure (black bars).
(d) Comparing the macroscopic properties of the empirical networks,
viz., their global efficiency $E$ and clustering $C$, with that of
degree-preserved randomized surrogates ($E^{(r)}$ and $C^{(r)}$, respectively) 
suggest that they are {\em infra-small world}.
Node colors represent distinct language families 
[Indo-European: English (en), Dutch (nl), Farsi (fa), 
French (fr), German (de), Russian (ru), Spanish (es) and Urdu (ur); Afro-Asiatic: Arabic (ar) and Hausa (ha); Austronesian: Malay (ms);
Turkic: Turkish (tr); Uralic: Finnish (fi)]. 
}
\label{fig1}
\end{figure}

A particular class of empirical 
networks that do not appear to fit into the regular/small-world/random 
(Rg/SW/Rn) spectrum 
are defined by the adjacent occurrence of characters in texts
of different natural languages which use alphabetic writing systems~\cite{note1}.
Fig.~\ref{fig1}~(a-b) shows that these networks are 
partially bipartite, 
comprising two clusters (consisting of vowels and consonants, respectively).
Most links occur between these two distinct types of nodes and 
comparatively few connect nodes of the same type. We find that these
networks have anti-modular character 
suggested by the block diagonal structure of their adjacency 
matrices {\bf A} ($A_{ij} = 1$ 
if nodes $i,j$ are connected, $=0$ otherwise)
with relatively sparsely populated diagonal blocks and dense
off-diagonal ones.
This is verified
by the negative values of the index $Q$, a quantitative measure for 
network modularity~\cite{Newman_spectra}, for the
empirical networks [Fig.~\ref{fig1}~(c)].
The macroscopic properties of these anti-modular networks 
show co-occurrence of extremely low clustering
(even lower than corresponding degree-preserved randomized networks) with 
high efficiency [Fig.~\ref{fig1}~(d)]. 
It is worth noting that in the usual Rg/SW/Rn paradigm the lowest
clustering and the highest efficiency that can be achieved correspond
to those of ER random networks, which form one of the extreme ends
of the small-world spectrum. 
However,
the empirical networks with anti-modular character have even lower
clustering and, in some cases, marginally higher efficiency, and 
are thus even ``smaller'' than the random graphs.
We thus term them {\em infra small-world}.

In this paper, we have shown that in general, networks comprising two
modules can be shown to have infra-small world character if the ratio
of inter- to intra-module connection density is varied so as to make them
anti-modular, with heterogeneity in module sizes making this
behavior very prominent.
In order to investigate the properties of such infra-SW
networks in a more detailed and systematic manner, we consider an ensemble 
of model networks. The topological organization
of the network connections can be tuned so as to change the mesoscopic
structure from modular to random and then to anti-modular (without 
altering the average degree $\langle k \rangle$ of the network) by gradually
increasing the density of inter- to intra-modular connectivity
$r = \rho_{out}/\rho_{in} \in [0, \infty)$~\cite{Pan09}. 
In order to consider heterogeneity in the size of the modules, we consider
that the $N$ nodes of each network are divided among two modules having
sizes $n$ and $N-n$, respectively. The size $n$ is randomly sampled 
from a Gaussian distribution with a mean of $N/2$ and whose sample standard deviation $\sigma_s$ is a free parameter 
quantifying the size heterogeneity~\cite{pathak}.
It is easy to show that the expected sizes of the two modules are $(N \pm 
\sqrt{2} \sigma_s)/2$.
By specifying $N$, $n$, $\langle k \rangle$ and $r$ estimated from 
empirical networks, we can construct corresponding model network ensembles
which have similar mesoscopic properties~\cite{note1}. 
Fig.~\ref{fig1}~(c) shows that the
model networks generated using parameters estimated from the 
orthographic networks of different languages can reproduce quantitatively the 
mesoscopic nature of the latter
fairly accurately.

\begin{figure}[!htb]
\centerline{\includegraphics[width=0.99\columnwidth]{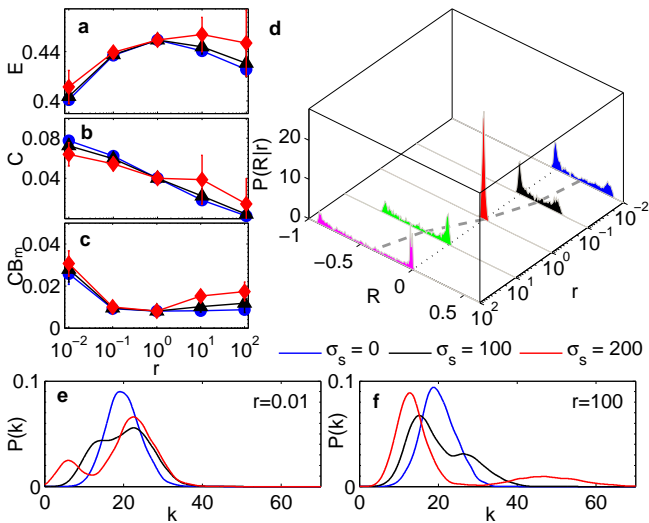}}
\caption{{\bf The macroscopic structural properties of a random network comprising two modules varies with its (anti-)modularity.} 
By increasing the ratio of inter- to intra-modular connection 
densities ($r$), the meso-scale nature of the
network  changes from being modular ($r<1$) to homogeneous ($r\simeq 1$) to anti-modular ($r>1$), while the standard deviation $\sigma_s$ in the number of elements belonging to each module quantifies the heterogeneity in module sizes.
(a-c) As the network changes from modular to homogeneous when $r$ is increased 
from $0$ to $1$, we observe an increase in the 
communication efficiency $E$ (a), 
while clustering $C$ (b) and 
maximum betweenness centrality ($CB_m$, c) decreases. 
However, with further increase of $r$ beyond $1$ the network becomes 
anti-modular and the structural properties now depend sensitively on 
module size heterogeneity. Thus, while $E$ decreases with $r$ when 
module sizes are comparable ($\sigma_s = 0$, represented by circles), 
when the sizes are very dissimilar (e.g., $\sigma_s = 200$, represented by diamonds) $E$ can increase to values higher than homogeneous random networks (a).
Similarly, higher module size heterogeneity, i.e., $\sigma_s \gg 0$ is related to reduction in the decrease of $C$ (b) and an increase in $CB_m$ (c) with increasing anti-modularity, compared to the $\sigma_s =0$ case.
(d) The distribution of the assortativity coefficient $R$ as a function of $r$ clearly indicates that modularity promotes assortative degree correlations while anti-modular networks are predominantly disassortative, the broken line indicating the variation of the mean value of $R$ with $r$. Note that, the distribution considers ensembles of networks with different size heterogeneities
($\sigma_s = 0, 50, 100, 150, 200$), with higher $\sigma_s$ resulting in greater deviation of a network from $R=0$ (neutral assortativity).  
(e-f) The dependence of the network properties on module size heterogeneity is related to the nature of the corresponding degree ($k$) distributions which develop a distinct bimodal character with increasing $\sigma_s$ in modular (e, $r=0.01$), as well as, anti-modular (f, $r=100$) scenarios.
For all cases shown here, we have used networks with $N=500$ and $\langle k \rangle =20$.
}
\label{fig2}
\end{figure}

To demonstrate that the infra-small world nature is associated with
anti-modular character of a network, we now characterize the model
networks 
in terms of their principal macroscopic properties.
Fig.~\ref{fig2}~(a) and (b) show 
the variation of the global efficiency $E$
and the average clustering coefficient $C$
as the mesoscopic structure of the
network is changed by varying $r$.
When the network is modular ($r < 1$), $E$ is lower and $C$ higher than the
corresponding values for homogeneous random networks ($r = 1$), consistent
with earlier observations that modular networks are small-world~\cite{Pan09}. 
On increasing $r$ beyond $1$, the network becomes anti-modular 
and approaches complete bipartivity as $r \rightarrow \infty$. This
makes connected triads increasingly unlikely, thereby decreasing
the clustering to values even lower than that of ER random networks. 
The efficiency of anti-modular networks, on the other hand, depend on the
extent of heterogeneity in module sizes. When module
sizes are similar (i.e., low $\sigma_s$) $E$ is seen to decrease 
monotonically from the maximum reached for $r=1$. However, for high $\sigma_s$,
when module sizes are very different, the efficiency attains values
even {\em higher} than that seen for the ER random networks.
This can be connected with the emergence of a bimodal degree
distribution in these networks with increasing $\sigma_s$ [Fig.~\ref{fig2}~(f)], 
with the smaller of the two modules comprising extremely high degree nodes. 
These
act as {\em hubs} connecting the entire network via extremely short paths
that pass through them, as indicated by the increased value of the maximum
betweenness centrality for such systems [Fig.~\ref{fig2}~(c)]. 
As mentioned earlier, these anti-modular networks having higher $E$ and 
lower $C$ compared to homogeneous random networks thus
lie beyond the spectrum of SW networks.
As it is known that higher efficiency enhances global
synchronization~\cite{Barahona2002} while high clustering hinders
it~\cite{McGraw2005}, such infra-SW networks have potential utility in 
applications where extremely rapid synchronization of activity
over the entire system (even faster than in random networks) is 
required. 

Further insight into the structural changes that the networks undergo as
$r$ is increased can be obtained by considering their degree homophily,
i.e., whether connected nodes have
a similar number of links, measured by
the Pearson correlation coefficient of degree between pairs of linked nodes,
$R$~\cite{Assor_Newman}.
Fig.~\ref{fig2}~(d) shows that when the networks are highly modular, they tend
to be degree assortative (i.e., $R>0$), particularly when module 
size heterogeneity is strong.
Indeed it is known that a network where high degree nodes belong to one
module while those of lower degree belong to another, exhibits
degree assortativity~\cite{Newman_spectra}.
This is consistent with the bimodal degree distribution for modular
networks shown in
Fig.~\ref{fig2}~(e), suggesting that the lower and higher peaks of 
the distribution correspond to the smaller and larger modules, respectively 
(see details below).
When $r$ is increased beyond $1$, making the networks anti-modular,
they become disassortative (i.e., $R<0$) when module sizes are unequal. 
These also have bimodal degree distribution [Fig.~\ref{fig2}~(f)], 
but with the
lower (higher) peak now associated with the larger (smaller) module.
The existence of disassortativity suggests that the anti-modular networks have 
star-like structures~\cite{NewmanGirvan2003}, with each high degree node of the 
smaller module preferring to connect to a large number of low degree
nodes in the bigger module. 

\begin{figure} [tbp]
\centerline{\includegraphics[width=0.9\columnwidth]{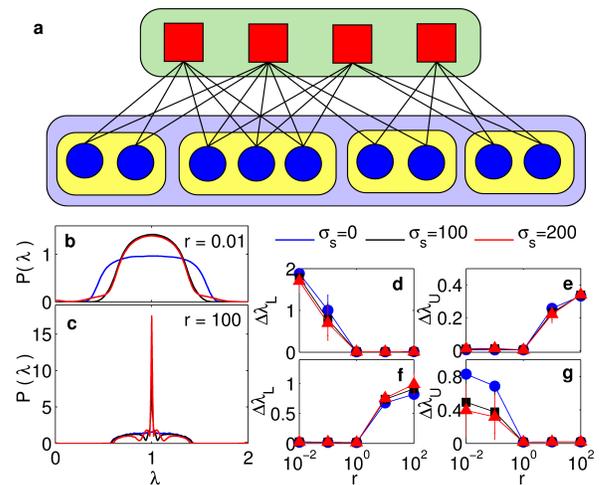}}
\caption{{\bf The variation of the spectral properties of a random network
comprising two modules with its (anti-)modularity.} As in
Fig.~\ref{fig2}, increasing $r$ changes the nature of the network from
modular ($r<1$) to anti-modular ($r>1$) while $\sigma_s$ quantifies the
heterogeneity in module sizes. (a) Schematic diagram of an anti-modular
network with two modules that are very heterogeneous in terms of size.
This simple system shows the genesis of bimodality in the degree
distribution, with nodes in the two modules having
 very different
number of connections. The nodes in the lower module are further subdivided 
into clusters (colored yellow) whose members are all connected to
exactly the same set of nodes. (b-c) While module size heterogeneity
can alter the eigenvalue distribution $P$($\lambda$) of the
corresponding normalized graph Laplacian $\mathscr{L}$ even for modular networks
(b, $r=0.01$), the occurrence of multiple nodes having identical
neighborhood for $\sigma_s \gg 0$ results in a prominent peak at
$\lambda =1$ for anti-modular networks (c, $r=100$).
The eigenvalue spectrum of both the normalized Laplacian $\mathscr{L}$ (d-e) and the modularity
matrix $B$ (f-g) exhibit systematic variation in the lower and upper
relative spectral gaps $\Delta \lambda_{L,U}$, respectively,
as functions of $r$ and $\sigma_s$. Eigenvalues are
arranged in increasing order, viz.,
$\lambda_{N-1} \geq \lambda_{N-2} \geq \ldots \geq \lambda_{2} \geq
\lambda_{1} \geq \lambda_0$ (for $\mathscr{L}$, $\lambda_0= 0$).
}
\label{fig3}
\end{figure}
For a network with two modules of unequal sizes $n, N-n$ and 
having overall mean degree $\langle k \rangle$, 
the average number of connections for the nodes in each of the two modules
can be very different, viz., $\langle k^{(n)} \rangle = \rho_i 
[n+r(N-n)]$ and $\langle k^{(N-n)} \rangle = \rho_i [(N-n)+rn]$, where
$\rho_i = \langle k \rangle/[N+2m(r-1)(1-\frac{n}{N})]$ is the
intra-module connection probability. For strong module
size heterogeneity, as the size of the networks become large 
(i.e., $n/N \rightarrow 0$ as $N\rightarrow \infty$),
$\langle k^{(n)} \rangle \sim \langle k \rangle r/[1+2(r-1)\frac{n}{N}]$
while $\langle k^{(N-n)} \rangle \sim \langle k \rangle (1+r\frac{n}{N})/
[1+2(r-1)\frac{n}{N}]$. From these expressions, it follows that in
modular networks ($r<1$), the larger of the two modules has a degree
distribution centered about a value close to $\langle k \rangle$ while
the nodal degrees of the smaller module approach $0$ as $r \rightarrow 0$.
On the other hand, for anti-modular networks, the average degree of the
larger module decreases asymptotically to $\langle k \rangle/2$ as the
network becomes completely bipartite, while 
that of the nodes in the smaller module initially increases linearly with
$r$ and eventually saturate to $\langle k \rangle N/2n$ as 
$r \rightarrow \infty$. This suggests that for strong module
size heterogeneity, the nodes in the smaller module will have 
degree $\gg \langle k \rangle$, making them hubs.
Thus, as the network organization changes
from modular to anti-modular, its topological structure alters from being
composed of two relatively weakly connected clusters, each having dense 
intra-connectivity, to one having a few hubs that tend to avoid
connecting to each other.
We note in passing that networks with bimodal degree distribution have
been shown to be dynamically more stable~\cite{Pan07,Brede_Sinha}, as well as, 
robust with respect to breakdowns and attacks~\cite{Tanizawa2005}.

\begin{figure}
\centerline{\includegraphics[width=0.9\columnwidth]{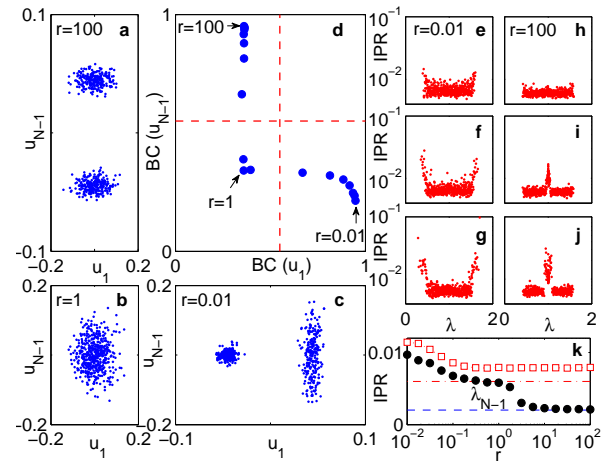}}
\caption{{\bf (Anti-)modularity is indicated by bimodal distribution in the
extreme eigenvectors of the Laplacian,
with a localization-delocalization
transition accompanying the change of the
meso-scale nature of the network.}
(a-c) The scatter plots of the eigenvector components corresponding
to the largest ($u_{N-1}$) and smallest non-zero ($u_1$) eigenmodes of the 
Laplacian matrix 
are shown as
the network changes from (a) anti-modular ($r=100$),
through (b) homogeneous ($r=1$) to (c) modular ($r=0.01$).
While modular networks are characterized by
a bimodal nature of the distribution of $u_{1}$ components, anti-modular
ones show a bimodal distribution for the components of $u_{N-1}$. 
This is quantified by the Bimodality Coefficient (BC) for each eigenmode.
As shown in panel (d), as $r$ is increased from $0.01$ (modular) to 
$100$ (anti-modular), the
position of the system in the \{BC($u_1$),BC($u_{N-1}$)\} plane moves
from  bottom right to top left
with the change in the meso-scale nature of the network.
The broken horizontal and vertical lines 
indicate the threshold value BC$^* = 5/9 \approx 0.555$ which corresponds
to a uniform distribution, with BC$\gg$BC$^*$ (BC$<$BC$^*$) suggesting 
that the distribution is bimodal (unimodal).
(e-j) Increased localization of Laplacian eigenvectors, measured by 
the inverse participation ratio (IPR) for the eigenmodes $\lambda$ 
is seen with greater diversity in the sizes of the two modules for 
both (e-g) modular ($r=0.01$) and (h-j) anti-modular 
networks ($r=100$) comprising two modules of sizes $n$ and $N-n$ 
[shown for $N=500$ and $\langle k \rangle =20$]. As the module size 
heterogeneity is increased from $\sigma_s=0$ (e,h) to $\sigma_s=50$ (f,i) 
and $100$ (g,j), we observe the emergence of prominent maxima in the 
IPR values. These occur close to the lowest (non-zero) and highest eigenvalues 
for modular networks [seen prominently in panels (f) and (g)] while, for anti-modular networks the peak appears around $\lambda=1$ [see panels (i) and (j)].
(k) A localization-delocalization transition for the principal eigenvector 
corresponding to the largest eigenvalue $\lambda_{N-1}$ (filled circles), whose IPR for
modular networks ($r\ll1$) is close to the maximum IPR value obtained for any 
Laplacian eigenmode (squares). As $r$ increases to $1$, the IPR becomes 
same as that expected for a random matrix (dash-dotted line). 
For $r \gg 1$, where the network is anti-modular, we
observe delocalization corresponding to 
IPR = $1/N$ (broken line).
}
\label{fig4}
\end{figure}
We can estimate the average path length $\ell$ (which is inversely related to
the global efficiency) of these networks as a function of their anti-modular
character and module size heterogeneity. For $r \geq 1$, assuming that
the local neighborhood of each node resembles a tree such that
cycles do not play a prominent role in the calculation of path lengths, one can 
use the approximation 
$[\langle k^{(n)} \rangle \langle k^{(N-n)} \rangle]^{\ell/2}
\sim N$. This yields 
\begin{equation}
\ell \sim \frac{\ln(N)}{\ln(\langle k \rangle (1+2(r-1)\frac{n}{N})^{-1}\sqrt{r(1+r\frac{n}{N})})},
\label{eq}
\end{equation}
which, further simplifies to $\ell \sim 2 \ln N/ \ln (N\langle k \rangle^2/4n)$
in the limit $r \rightarrow \infty$ (i.e., when the network becomes completely
bipartite. It is easy to see that as $n/N$ is extremely small, the effective
path length reduces to values lower than the equivalent ER network ($\ell_{rand}
\simeq \ln N/\ln \langle k \rangle$), providing the basis for the
infra-small world property of anti-modular networks. 
Note that, this implies that spreading processes will be much faster on
partially bipartite networks than even random ones. This is important
to consider, e.g., for epidemic propagation in 
livestock populations whose transport between farms and markets form
a partially bipartite network~\cite{Kiss2006}.

As mentioned earlier, strong module size heterogeneity in the anti-modular regime
results in the formation of star-like structures with nodes of the smaller module
acting as hubs. As many of the nodes in the larger module connect to the 
same set of hubs, we can cluster them into groups of nodes having identical
neighborhoods~[Fig.~\ref{fig3}~(a)]. The occurrence of multiple nodes that have
exactly the same neighbors is reflected in the degeneracy of the
unity eigenvalues of the
normalized symmetric Laplacian matrix $\mathscr{L} =${\bf I}$-${\bf D}$^{-1/2}${\bf A}{\bf D}$^{-1/2}$, where {\bf I} is identity matrix and {\bf D} 
is a diagonal matrix with $D_{ii}=k_i$, i.e., the degree of $i^{th}$ node,
for the network~\cite{Chung2003}. 
This can be seen in the prominent peak at $\lambda=1$ 
seen for sufficiently strong heterogeneity, e.g., curves corresponding to
$\sigma_s>0$ in Fig~\ref{fig3}~(c). This contrasts with the 
spectral behavior of 
$\mathscr{L}$ for modular networks (i.e., $r<1$) shown in Fig.~\ref{fig3}~(b),
as well as, with the semi-circle law
expected for ER random networks~\cite{Chung2003}. 
We note that when the module sizes are similar,
eigenvalue distributions for $\mathscr{L}$ of
modular, as well as, anti-modular networks are platykurtic in nature
(indicated by the excess kurtosis of the bulk of the distribution being $-1$)
as is also the case for ER random networks.
However, heterogeneity in module sizes leads to very different behavior 
for networks with $r<1$ and $r>1$.
For modular networks, increasing $\sigma_s$ above $0$ initially raises the
excess kurtosis of the $\lambda$ distribution to $0$.
However, further increase of heterogeneity results in the larger module to 
dominate the system properties and the semi-circle law
is recovered for the bulk 
in the limit of large $\sigma_s$.
However, for anti-modular networks, increasing $\sigma_s$ makes the 
excess kurtosis positive, suggesting that the eigenvalue distribution 
becomes leptokurtic in the presence of strong module size heterogeneity.

Another important spectral characteristic of $\mathscr{L}$ that is
intimately related to the meso-level structural organization of the
network is the relative size of the gaps occurring at the lower and upper ends
of the eigenvalue spectrum, viz., $\Delta \lambda_L =
2 (\lambda_2^{-1}-\lambda_1^{-1})/(\lambda_2^{-1} +\lambda_1^{-1})$ and
$\Delta \lambda_U = 2(\lambda_{N-1}^{-1}-\lambda_{N-2}^{-1})/
(\lambda_{N-1}^{-1}+\lambda_{N+2}^{-1})$,
respectively. As the network becomes strongly modular ($r \ll 1$), 
there is a corresponding decrease in the smallest non-zero
eigenvalue $\lambda_1$ of $\mathscr{L}$ (in the limit $r\rightarrow 0$,
$\lambda_1=0$).
As a result, the lower spectral gap is seen to increase with decreasing
$r$ [Fig.~\ref{fig3}~(d)], which is associated
with the appearance of distinct time-scales for the dynamics
occurring at different scales in the network, viz., fast intra-modular
and slower inter-modular processes~\cite{Pan09}.
Conversely, when the anti-modular
character becomes more prominent as $r \rightarrow \infty$, the
largest eigenvalue $\lambda_{N-1}$ of $\mathscr{L}$ approaches its
maximum value ($=2$) and the upper spectral gap is seen to
increase [Fig.~\ref{fig3}~(e)]. Such an association between the
mesoscopic structural organization of the network and its spectral
characteristics is also reflected in the
corresponding gaps of the eigenvalue spectrum for the modularity matrix
{\bf B} [{$B_{ij} = A_{ij} - (k_i k_j/2l)$, where $l$ is
the total number of connections}]~\cite{Newman_spectra}.
We observe that the spectral gaps for {\bf B} are more sensitive
to heterogeneity in module sizes than the corresponding quantities for
$\mathscr{L}$. Large differences in the
sizes of the two modules can mask the modular character of a network
(for $r<1$)
as the larger module dominates the system (indeed
$Q$ is seen to decrease with increasing heterogeneity).
Hence, with increasing $\sigma_s$,
we find that the upper spectral gap of {\bf B}, which
is linked to modularity,
decreases [Fig.~\ref{fig3}~(g)].
However, for $r>1$,
increasing $\sigma_s$ will make the
distinct identity of the nodes belonging to the two ``modules''
even more prominent in terms of their degree (the limiting case corresponding
to star-like networks). As a result, the lower spectral gap, which
contributes to information about the anti-modular character of the network,
increases with $\sigma_s$ [Fig.~\ref{fig3}~(f)].

Focusing now on the properties of the eigenvectors of ${\mathscr L}$,
we observe that the eigenmodes corresponding to $\lambda_1$ and
$\lambda_{N-1}$ convey information about the two modules into
which the network is partitioned. Thus, for modular networks ($r<1$),
the group to which each node belongs can be identified from the
sign of the corresponding component of $u_1$, the eigenvector
associated with $\lambda_1$. On the other hand, for anti-modular
networks ($r>1$), this role is played by $u_{N-1}$, the
eigenvector corresponding to the largest eigenvalue $\lambda_{N-1}$.
Specifically, the distribution of the eigenvector components shows
a bimodal nature, which becomes more prominent as
$r$ approaches $0$ (for $u_1$) or diverges (for $u_{N-1}$) [see panels (c)
and (a), respectively, of Fig.~\ref{fig4}]~\cite{note1}.
For a homogeneous ER random network ($r=0$), where such a partitioning is
not possible, the distributions
for both of these eigenvectors are unimodal [Fig.~\ref{fig4}~(b)].
These observations suggest that we can identify the existence of 
anti-modular mesoscopic organization in a network by
measuring the extent of bimodality in the
distribution of components for $u_{N-1}$. 
For this purpose, we calculate the Bimodality Coefficient 
$BC = (m_3^2 +1)/(m_4 + 3 [(n-1)^2]/(n-2)(n-3)])$, where
$m_3$ is skewness of the
distribution and $m_4$ is its excess kurtosis~\cite{Pfister2013}.
Fig.~\ref{fig4}~(d) shows how model networks corresponding to
different values of $r$ can be characterized in terms of $BC$s of
$u_1$ and $u_{N-1}$. Thus, modular networks ($r<1$) are
characterized by strong bimodality in $u_1$ with $BC(u_1) \gg BC^*$,
where $BC^*=5/9$ corresponds to uniform distribution, while anti-modular
networks ($r>1$) are seen when $u_{N-1}$ has strong bimodality.

The eigenvectors of $\mathscr{L}$ also exhibit
localization behavior associated with structural heterogeneities that 
can inform us about the outcome of diffusive processes
on a network~\cite{Nakao2010,Hata2017}. We quantify the
localization of the $k$th eigenvector $u_k$ by its inverse participation 
ratio, $IPR (u_k) = \sum^N_{i=1} [u_{k,i}^4]$ where $u_{k,i}$ are the components
of the eigenvector~\cite{Distint_EigVec_loc,IPR_Phys_Lett}. Complete
delocalization is associated with the minimum value of $IPR (= 1/N)$, 
when all components have equal contribution. Conversely, the maximum
value of $1$ is associated with extreme localization, obtained when an 
eigenvector has only a single 
component having a finite contribution.
Fig.~\ref{fig4}~(e-g) show that a modular network exhibits high 
values of IPR for eigenmodes at both the lower and higher ends
of the eigenvalue spectrum. On the other hand, anti-modular
networks show delocalization in the principal eigenmodes while 
having strong localization in the central modes [Fig.~\ref{fig4}(h-j)]. 
Localization in both types of networks becomes more prominent with increasing 
module size heterogeneity $\sigma_s$.
The very different nature of localization behavior in modular
and anti-modular networks is reflected in the localization-delocalization 
transition seen for the principal eigenmode (associated
with the largest eigenvalue $\lambda_{N-1}$ of $\mathscr{L}$)
as $r$ is varied [Fig~\ref{fig4}~(k)]. Thus, as the mesoscopic nature 
of the network changes from modular to anti-modular, we observe that
the eigenmode becomes completely delocalized ($IPR \rightarrow 1/N$ as
$r$ diverges), irrespective of the extent of heterogeneity
in module sizes (similar to transitions seen in the spectral behavior
of network adjacency matrices~\cite{Slanina2017}).

To conclude, we have shown that networks having a partially bipartite
structure exhibit properties that place them outside the well-known
Rg/SW/Rn range of network structures. In particular,
when the sizes of the two partitions into which the nodes are grouped
are very unequal, the network has a communication efficiency higher 
than that of homogeneous ER random networks, and correspondingly
lower clustering. Such infra-SW property is
related to the anti-modular character of these networks which we
demonstrate by analyzing an ensemble of model networks whose
mesoscopic nature can be systematically varied.
Our work also suggests signatures, such as BC for the
principal eigenvector of the corresponding Laplacian, to identify
potential anti-modular organization in a wide range
of empirical networks.
We observe that for strong
module size heterogeneity, the degree distribution of anti-modular networks 
becomes bimodal, which can make such network robust against a variety
of perturbations. As anti-modular structure has been reported in
several empirical situations, such as, for
networks representing the adjacent occurrence of different parts 
of speech~\cite{Newman_spectra}, bilateral investment agreements between
nations~\cite{Saban2010}, romantic online interactions~\cite{Holme2003},
food webs~\cite{Townsend1998,Estrada2005} and that of farms and markets connected
by movement of livestock~\cite{Kiss2006},
it is important to understand whether such an organization appears
because of functional considerations. Understanding how 
relative contributions of intra- and inter-dependence in networks
comprising multiple partitions can impact, for instance, their
robustness~\cite{Singh2019}, will be a challenging problem 
for the future.

\begin{acknowledgments}
This work was supported in part by IMSc Project of Interdisciplinary 
Science \& Modeling (PRISM) and IMSc Complex Systems Project (XII Plan) 
funded by the Department of Atomic Energy, 
Government of India.
We would like to thank Shakti Menon and K Chandrashekar
for helpful discussions.
The simulations required for this work were done in the
Nandadevi cluster of the IMSc HPC facility.
\end{acknowledgments}


\onecolumngrid
\vspace{20cm}
\section*{\large{Supplementary Material\\ for\\
Anti-modular nature of partially bipartite networks makes
them {\em infra small-world}}}

\begin{center}
	{Aradhana Singh$^{1}$, Md. Izhar Ashraf$^{1,2}$ and
	Sitabhra Sinha$^{1,3}$}\\
	\bigskip
	{\em \small $^1$The Institute of Mathematical Sciences, CIT Campus,
	Taramani, Chennai 600113, India.\\
	$^2$ BS Abdur Rahman Crescent Institute of Science \& Technology, 
	Vandalur, Chennai 600048, India.\\
	$^3$Homi Bhabha National Institute, Anushaktinagar, Mumbai 400094,
	India.\\}
\end{center}
\setcounter{figure}{0}
\setcounter{equation}{0}
\renewcommand\thefigure{S\arabic{figure}}
\renewcommand\thetable{S\arabic{table}}
\renewcommand{\thesection}{\Roman{section}} 
\renewcommand{\thesubsection}{\thesection.\Alph{subsection}}

\section{Data Description}
For construction of the empirical networks defined by the  adjacent occurrence 
of characters in texts of different natural languages that are
written using alphabetic systems (shown in Figure 1 of the main text), 
we have used several different sources that are described below.

\subsection{Phoneme Network:}
In order to construct the network connecting adjacent phonemes
[Figure 1(a)] that occur in English words we have used a subset
of 5321 words from a lemmatized
 list of 6318 frequently used words 
(i.e., with $>800$ occurrences) from the British National Corpus 
(\url{https://www.kilgarriff.co.uk/BNClists/lemma.num}, 
accessed: 26th May 2016) which were phonetically transcribed using
the open-source Carnegie Mellon University Pronouncing 
Dictionary (\url{http://svn.code.sf.net/p/cmusphinx/code/trunk/cmudict/sphinxdict/cmudict_SPHINX_40}, 
accessed: 4th January 2018).
The phonetic output is given in terms of the 
ARPAbet phoneme set (\url{http://svn.code.sf.net/p/cmusphinx/code/trunk/cmudict/sphinxdict/SphinxPhones_40}, accessed: 4th January 2018). The ARPAbet is a standardized set
of 39 phonemes used for describing the pronunciation of words in different
languages. 
The phonetic transcription is subsequently mapped to the International Phonetic Alphabet (IPA) notation using
the mapping provided in \url{https://en.wikipedia.org/wiki/ARPABET}. The nodes of the network shown in Fig.~1~(a) are
labeled using the IPA symbols.

\noindent
\subsection{Orthographic Networks:}
For the networks described by the adjacent occurrence of alphabetic characters
in words written in different natural languages, we have used the following
language corpora:\\

{\bf Arabic:} We have used a database of 14867 unique words of Classical (or Quranic) Arabic, a Semitic language which was originally written using a 
consonantal alphabet (also known as an `abjad'). The present alphabet, considered an ``impure abjad'', comprises 27 signs representing
consonantal sounds (including a modifier and a glottal stop) and 9
signs that represent long vowels (3), as well as, combinations of long vowels with diacritical marks (3), 
diphthong (1)
and glottal stop (2). The database is created by selecting all words written using at least two alphabetic characters 
from {\em Tanzil}, an international project started in 2007
to produce a standard Unicode text for the Qur'an
(\url{http://tanzil.net/download/}, accessed: 25th March 2015).\\

\noindent
{\bf Dutch:} We have used 
$9146$ unique non-hyphenated words having
two or more characters from a list of the 10000 most commonly used words
in Dutch, a member of the Germanic branch of the Indo-European
language family. The data has been collected
from the {\em Wortschatz} website maintained by the University of
Leipzig
(\url{http://wortschatz.uni-leipzig.de/Papers/top10000nl.txt},
accessed: 22nd May 2015). The Dutch signary consists of
31
distinct alphabetic characters comprising 21 consonants, 5
vowels, 3 vowels with diacritical
marks (acute accents or diaeresis), the digraph `ij' that is
considered as a letter in the Dutch language
and
an extra letter from the German alphabet (the {\em Eszett}).\\

\noindent
{\bf English:} 
We have used the \textit{Mieliestronk} list of 58109
distinct words (comprising two or more letters) in English - belonging to the Germanic branch of the Indo-European language family - that has been compiled by merging several different word-lists
(\url{ http://www.mieliestronk.com/wordlist.html}, accessed: 4th
December 2011). 
The English signary is made up of 26 lower case letters of the English alphabet, comprising 5 vowels and 21 consonants. 
The list we have considered excludes spellings that are considered to be non-British. 
A hyphenated word is listed in unhyphenated form by removing the punctuation mark. The list contains singular and plural forms of several words, as well as, multiword phrases that are in common usage rendered as a single word. \newline

\noindent
{\bf Finnish:} A list of the 10000 most commonly
used words (all of which use two or more letters)
in the Finnish language, belonging to the Finnic branch of the
Uralic language family, has been used.
The data, obtained from the \textit{Wikiverb} website, has been
collected from
newsgroup discussions, press and modern literature
(\url{http://wiki.verbix.com/Documents/WordfrequencyFi}, accessed:
24th June 2015).
The Finnish signary has 25 distinct signs - i.e., all vowels and
consonants of the modern Latin alphabet along with two additional vowels ``\"{a}'' and ``\"{o}'', excepting ``q'',``x'' and
``w''.\\

\noindent
{\bf French:}  We
have chosen
9189 
unique words that are written using two or more alphabetic characters from a list of the 10000 most commonly used
words in French, a Romance language belonging to the Indo-European
family. The data
has been collected from the \textit{Wortschatz} website
maintained by the University of Leipzig
(\url{http://wortschatz.uni-leipzig.de/Papers/top10000fr.txt},
accessed: May 22nd 2015). The French signary has 30 distinct alphabetic
characters comprising 26 letters of the Latin alphabet along with 3
vowels with diacritical marks (acute accents or diaeresis)
and an apostrophe sign.
\\

\noindent
{\bf German:} We have chosen 
9152 distinct words that are represented using
two or more alphabetic characters from a list of the 9172 most commonly used
words in German, a member of the Germanic branch of the Indo-European
language family. The data
has been collected from the \textit{Wortschatz} website
maintained by the University of Leipzig
(\url{http://wortschatz.uni-leipzig.de/
Papers/top10000de.txt},
accessed: May 22nd 2015). The German signary has 32 distinct alphabetic
characters comprising the 26 letters of the Latin alphabet
along with 4 vowels having diacritical marks (umlauts or acute
accents), a ligature (the {\em Eszett} or {\em scharfes S}) and
an apostrophe sign.\\

\noindent
{\bf Hausa (Boko):} We have used a list of 
7062 unique
words that are written using two or more alphabetic characters, obtained from a Hausa online
dictionary maintained by the University of Vienna
(\url{http://www.univie.ac.at/
Hausa/KamusTDC/CD-ROMHausa/KamusTDC/ARBEIT2.txt},
accessed: 19th May, 2015).
The Hausa signary has 30 distinct alphabetic characters comprising 23
letters from the Latin alphabet, four additional signs
representing glottalized consonants, two digraphs (`sh' and `ts')
and an apostrophe sign.
\\

\noindent
{\bf Malay (Rumi):} We have chosen 9970 unique words that
are written using two or more alphabetic characters from a list of 10000 most commonly
used words in Malay, a member of the Austronesian language family. 
All the words are written in {\em Rumi}
or Latin script, which is the most commonly used form for writing
Malay at present, although a modified Arabic script ({\em Jawi}) also
exists.
The data has been collected from the list of high frequency words
that are publicly available at {\em Invoke IT Blog}
(\url{https://invokeit.wordpress.com/frequency-word-lists/},
accessed: 4th January, 2014).
The signary comprises the 26 letters of the Latin alphabet.
\\

\noindent
{\bf Persian:} We have used a list of 10000 most commonly used words
(each represented using two or more characters) in Persian, a member of
the Indo-Iranian branch of the Indo-European language family, which is
written using a modified form of the consonantal Arabic alphabet (an `abjad'). The
words are obtained from a list of high-frequency words compiled using
the Tehran University for Persian Language corpus and available at
{\em Invoke IT Blog}
%
(\url{https://invokeit
.wordpress.com/frequency-word-lists/},
accessed: 4th January 2014).
The signary comprises 40 signs, viz., 32 consonantal signs, a long
vowel indicator (`alef madde'), a ligature
(`l\={a}m alef'), a diacritic (`hamze'), 3 consonants with the
`hamze' diacritical mark and different forms
for the consonants `k\^{a}f' and `ye' when they occur in final
position.
\\

\noindent
{\bf Russian:} We have used a list of 
9011 distinct words that are written using two
or more alphabetic characters in Russian, a member of the Slavic branch of the
Indo-European language family, written using a Cyrillic
alphabet. The data has been collected from
{\em Russian Learners' Dictionary: 10,000 words in frequency order}
compiled by Nicholas J Brown (Routledge, London, 1996), after removing
all words that use characters which are not included in the standard Russian alphabet.
(\url{https://docs.google.com/spreadsheets/d/1hSsPR0fN7I456-TZOUFJwOb7GjSrqeoOo02hMCy9NfI/edit?pli\=1\#gid\=7},
accessed: 18th May 2015),
The signary comprises the 33 letters of the modern Russian alphabet, comprising 10 vowels, 21 consonants and 2 signs
that indicate pronunciation.
\\

\noindent
{\bf Spanish:} We have used a list of  
4902 distinct high-frequency words (that are written using two or more alphabetic characters) in
Spanish, a Romance language belonging to the Indo-European family.
The data has been collected from {\em A Frequency Dictionary of
Spanish} compiled by Mark Davies (Routledge, New York, 2006).
The Spanish signary uses 35 distinct alphabetic characters comprising 26
letters of the basic Latin alphabet along with an additional character
\~{n} and two digraphs (`ch' and `ll'), as well as, vowels with
diacritical
marks (acute accents or diaeresis).\\

\noindent
{\bf Turkish:} We have used a list of 
9909 distinct high-frequency words (that
are written using
two or more alphabetic characters) in Turkish, a member of the Turkic language
family. The data has been collected from a {\em Wiktionary} word
frequency list
(\url{https://en.wiktionary.org/wiki/
Wiktionary:Frequency\_lists/
Turkish\_WordList\_10K}, accessed: 14th July 2015).
The signary used has 32 letters, comprising 29 letters of the Turkish
alphabet and 3 vowels used in conjunction with circumflex
accents.\\

\noindent
{\bf Urdu:} We have used a database of
4998 unique words (that are represented using two or more characters)
in Urdu, an Indo-Aryan language belonging to the Indo-European family,
that is written using an extended Persian alphabet. The words are
obtained from a list of frequently used words maintained by the
Center for Language Engineering at Lahore
(\url{http://www.cle.org.pk
/software/ling\_resources/UrduHighFreqWords.htm},
accessed: 1st January 2014).
The signary comprises 46 signs, viz., 35 consonantal signs and 11 signs that represent long vowels  (4), vowels  with diacritics  (2), vowels 
used in conjunction with a glottal stop (2), a diphthong (1) and two additional signs used for writing certain loan-words  (2).

\section {Construction of Adjacency Matrix from Empirical Data} 
\begin{figure} [tbp]
\begin{center}
\includegraphics[width=0.9\linewidth]{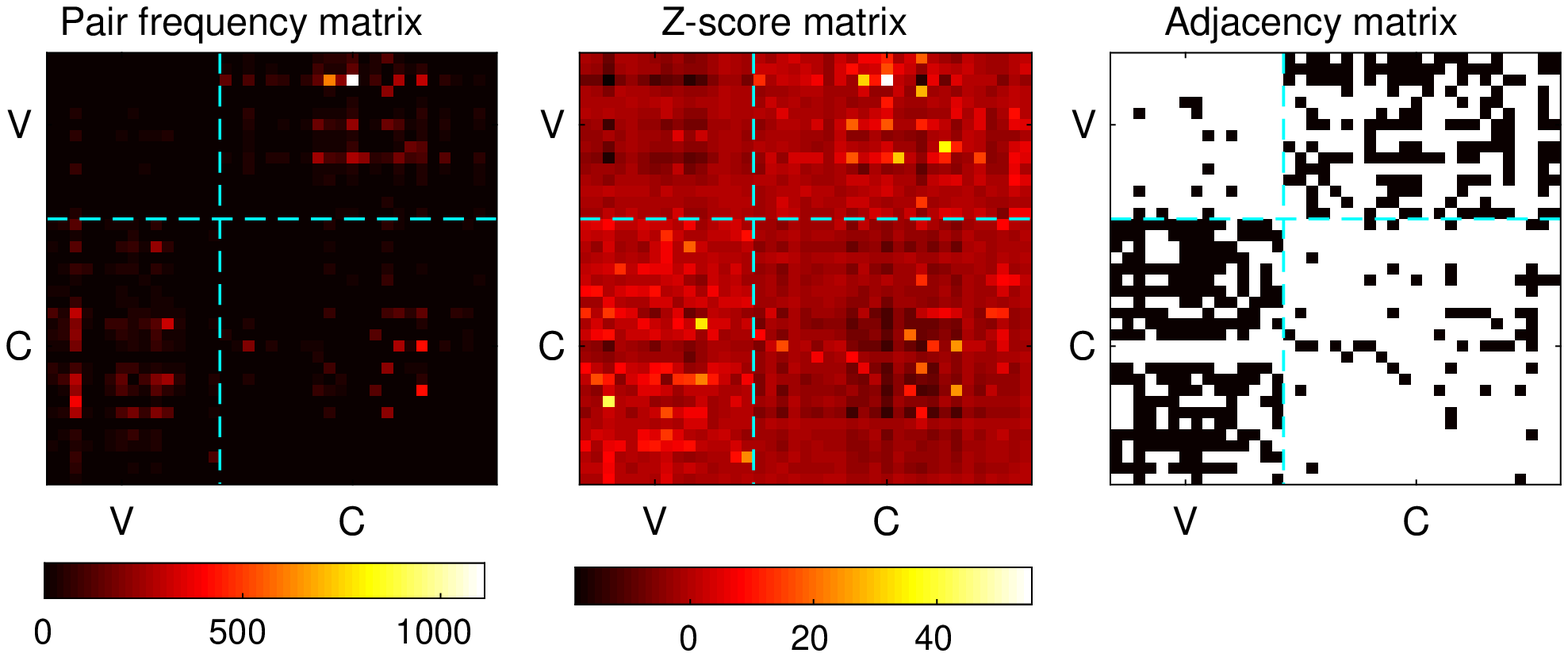}
\end{center}
\caption{Schematic diagram indicating 
the process of network representation of statistically significant occurrence of phonemic or alphabetic signs in adjacent positions in a natural
language corpus. The three matrices shown are successively generated, beginning with (left) one showing co-occurring sign pair frequencies obtained from 
the corpus. The second (middle) quantifies the statistical significance (quantified in terms of $Z$-score) of such co-occurrences by comparing the empirical frequency against that obtained from $10^3$ random surrogates, each constructed by randomly permuting the graphemes of every word in the corpus. Finally, an adjacency matrix (right) is obtained by imposing a threshold on the $Z$-score, connecting two signs if the frequency of co-occurrence in adjacent positions is higher than that expected by chance.}
\label{figs1}
\end{figure}

In order to construct the networks representing adjacent occurrence of graphemes in written texts, we have considered distinct phonemes
(for phoneme network) or alphabetic signs (for orthographic network) as the nodes of the network. Connections between two nodes
are made based on statistically significant co-occurrence of the two graphemes, corresponding to the two nodes, in adjacent
positions in words included in the corpus under consideration. For example, consider two graphemes $x_1$ and $x_2$ that occur in a particular
corpus. Let $n(x_1,x_2)$ be the number of times they are found in adjacent positions in the words that occur in the database.
We need to compare this with the frequency of co-occurrence entirely by chance. This is computed from a random surrogate
of the database, which is constructed by randomly permuting the graphemes in every word of the original database. 
From $M$ such realizations of random surrogates, we obtain the mean $\langle n^{rand}(x_1,x_2) \rangle$ and
standard deviation $\sigma^{rand} (x_1,x_2)$ of the frequency with which $x_1$ appears next to $x_2$ in a word simply 
as a chance outcome of their respective total frequencies of occurrence in the entire database. For the databases
considered here, we have used $M=10^3$.
Thus, we can define a measure of the statistical significance of the empirical frequency $n(x_1,x_2)$ as
\begin{equation}
Z(x_1,x_2) = \frac{n(x_1,x_2) - \langle n^{rand}(x_1,x_2)\rangle}{\sigma^{rand}(x_1,x_2)} .
\end{equation}
If $Z>0$ for any pair of graphemes, it suggests a possible significant association between them as they co-occur more than what
is expected by chance. Therefore, by assigning a link between two nodes $i$ and $j$ whenever the $Z$-score for the pair of graphemes
$x_i$ and $x_j$ associated with these is positive, we can define a network represented by the   
adjacency matrix {\bf A}, where $A_{ij} = 1$ if $Z(x_i,x_j)>0$ and $ = 0$, otherwise (Fig.~\ref{figs1}).
We note that, in general, $A_{ij} \neq  A_{ji}$, as the frequencies of adjacent occurrence of two graphemes are different depending on
the order in which they occur in words, i.e., $n(x_i,x_j) \neq n(x_j,x_i)$.\\

\begin{figure} [tbp]
\begin{center}
\includegraphics[width=0.99\linewidth]{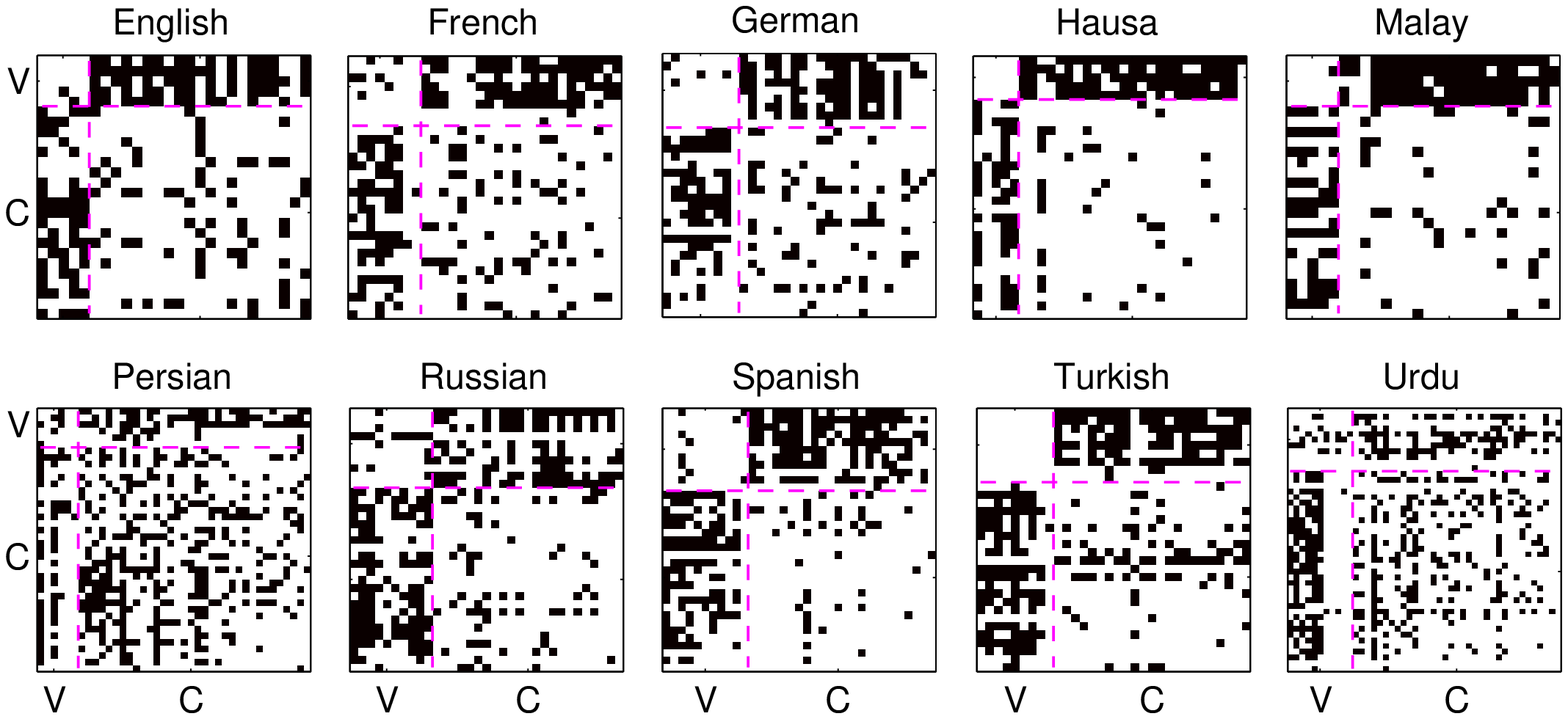}
\end{center}
\caption{Adjacency matrices representing orthographic networks, obtained by connecting alphabetic characters that occur significantly often at consecutive positions in words in different natural langugages, viz., English, French, German, Hausa, Malay, Persian, Russian, Spanish, Turkish and Urdu. Filled black squares at any position in the matrix (indicating $A_{ij} = 1$) represent the existence of a directed link between the pair of corresponding alphabetic characters
(viz., graphemes $x_i$ and $x_j$), whose co-occurrence frequency is more  than what is expected by chance. As in the case of the adjacency
matrices for Arabic, Dutch and Finnish shown in Fig.~1(b) in the main text, these also exhibit relatively higher density of connections between the
two modules comprising vowels (V) and consonants (C) respectively, as compared to connections within each module. It suggests anti-modular organization of the orthographic networks for a variety of languages that use an alphabetic writing system. See Table~\ref{tabs1} for 
details on the number of alphabetic signs of each type (vowels and consonants) and the topological properties of the network for the
different languages.
}
\label{figs2}
\end{figure}
\begin{table} [tbp]
\begin{center}
\includegraphics[width=0.9\linewidth]{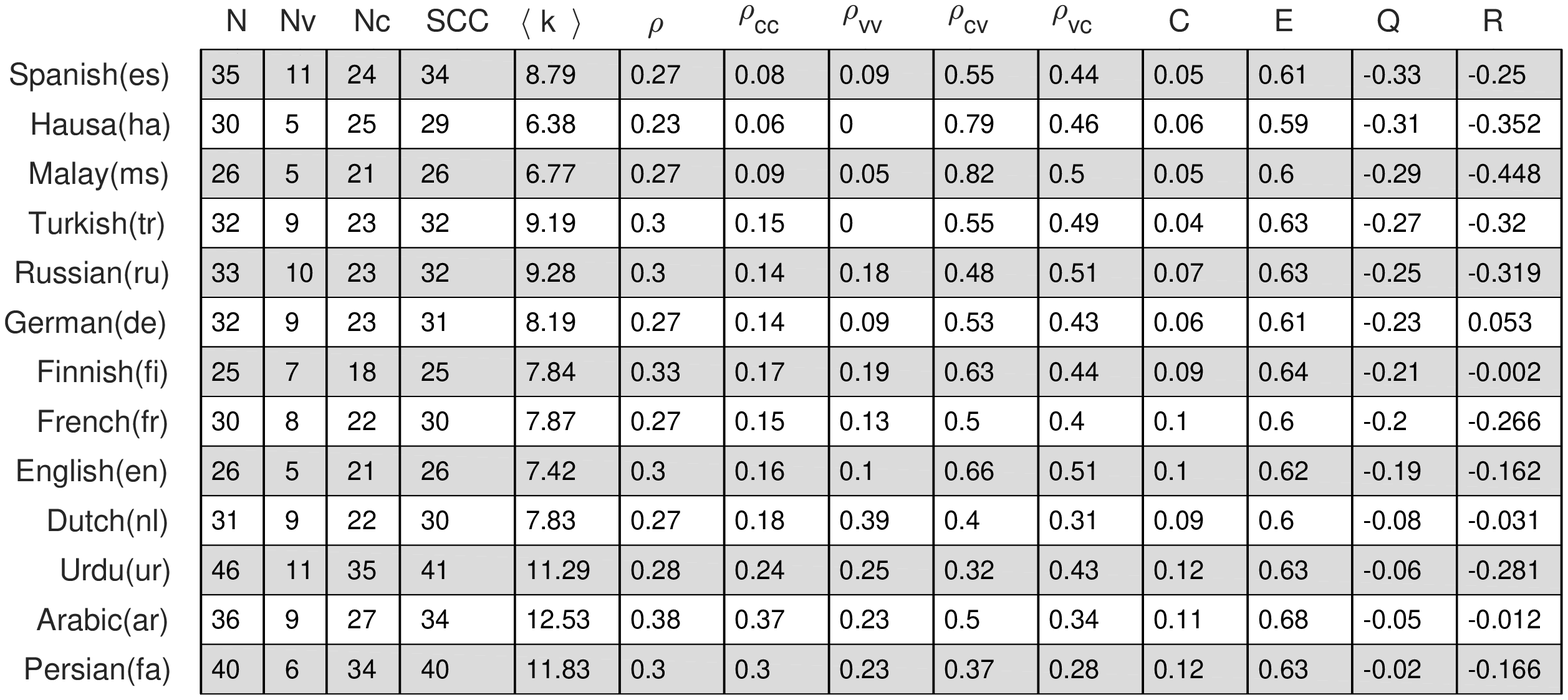}
\end{center}
\caption{Properties of the orthographic networks constructed from corpora of different natural languages: $N$ is signary size, $Nv$ \& $Nc$ are the number of  graphemes representing vowels and consonants, respectively, SCC is the size of the strongly connected component of the network, $\langle k\rangle$ is the average degree, $\rho$ is the connection density, $\rho_{cc}$ \& $\rho_{vv}$  are intra-modular connection densities (within
consonants and within vowels, respectively) while $\rho_{cv}$ \& $\rho_{vc}$ are inter-modular connection densities, $C$ is the clustering coefficient, $E$ is the communication efficiency, $Q$ is the modularity index and $R$ is the assortativity coefficient of the network. The standard abbereviations
for the names of the different languages, corresponding to each row, are indicated in parentheses. 
}
\label{tabs1}
\end{table}
\noindent 
In the main text, the adjacency matrices constructed using the above procedure for Arabic, Dutch and Finnish have been shown [Fig.~1(b)].
Fig.~\ref{figs2} shows the adjacency matrices for ten other languages. Table~\ref{tabs1} provides detailed information about each of these
orthographic networks. Apart from mentioning the total number of nodes (corresponding to different graphemes) and the number of vowels
and consonants (or, rather non-vowels) which provide the sizes of the two partitions into which the nodes are divided, the different columns
indicate the size of largest connected component (i.e., the set of nodes for which a directed path exists from any node to any other node),
the average number of connections per node, the overall connection density as well as the density within the two compartments and between
two compartments, and network metrics such as the average clustering coefficient, the communication efficiency, the modularity index and
the assortativity coefficient. \\

\begin{figure}
\begin{center}
\includegraphics[width=0.99\linewidth]{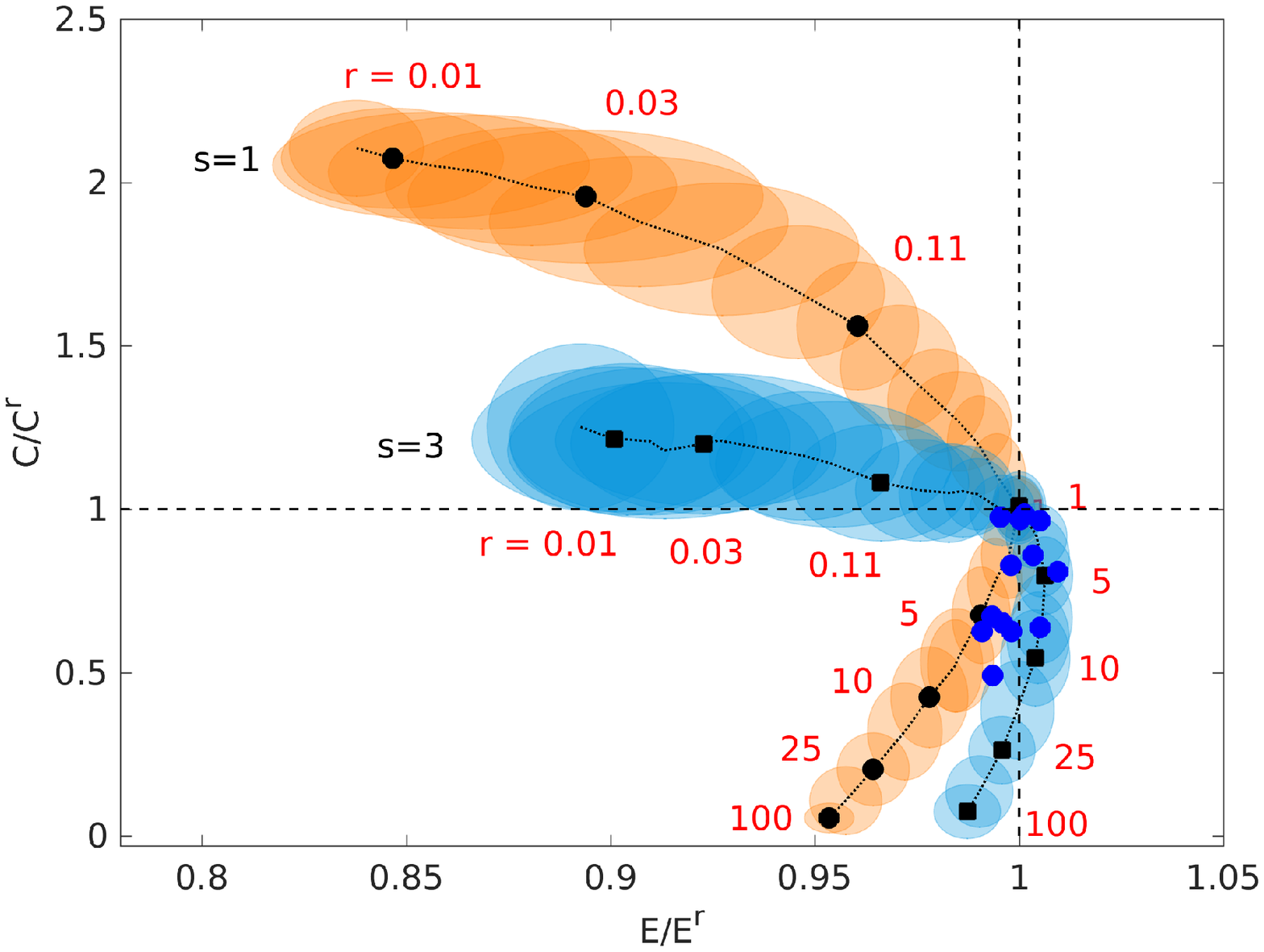}
\end{center}
\caption{Variation of the macroscopic properties, viz., global communication efficiency $E$ and mean clustering coefficient $C$,  of model
networks (consisting of two modules) as a function of the ratio of the inter- to intra-modular connection densities, $r$. Values of both the network
metrics $E$ and $C$ are expressed relative to the corresponding values for degree-preserved randomized networks, viz., $E^r$ and $C^r$.
Role of module size heterogeneity can be inferred by comparing the situation where modules are of equal size ($s=1$,black circles) with
the case where one module is about three times larger than the other ($s=3$,black squares). Each value is obtained by averaging over 50
realizations of a given model network, while colored ovals around each symbol represent the standard deviations for the metrics 
calculated over the realizations. For comparison, the values for the empirical
orthographic networks [shown in Fig.~1(d) in the main text] are also indicated (blue circles).  
Note that the total number of nodes $N$ ($=32$) and average degree $\langle k \rangle$ ($= 8$) of the model networks have been kept
constant for all $r$ and $s$ considered, in order to be comparable to that of the empirical networks.  
}
\label{figs3}
\end{figure}

\noindent
To show the infra-modular nature of the anti-modular networks, we have compared their global properties, specifically, their average
clustering coefficient $C$ and communication efficiency $E$, with the corresponding quantities $C^r$ and $E^r$ of the randomized network 
counterparts that have the same degree sequence as the anti-modular networks. Fig.~\ref{figs3} shows how these two network
metrics vary (relative to those of randomized networks) as we change the mesoscopic nature of the model networks from
modular to anti-modular. This is done by systematically increasing the ratio of inter- to intra-modular connection density, $r$, from
values less than $1$ (when the network is modular) to values greater than $1$ (when the network becomes anti-modular).
We have also shown the effect of module size heterogeneity by contrasting the situation where the module sizes are same with one
where they are different. To quantify the heterogeneity we have used the ratio of the size of the larger to the smaller partition.
If $N$ be the total number of nodes and $n$ is the number of nodes in the larger partition, then this ratio corresponds to
$s = n/(N-n)$. The two situations we consider are $s=1$ (i.e., where the module sizes are same) and $s=3$. As can be
seen from Fig.~\ref{figs3}, anti-modular networks, particularly in the presence of appreciable module size heterogeneity, exhibit higher
communication efficiency and lower clustering than their randomized network counterparts.\\

We have also investigated the spectral properties of the networks (which are directed, in general), focusing on the normalized symmetric Laplacian matrix $\mathscr{L}$ defined for the strongly connected component of the network as follows:
\begin{equation}
\mathscr{L}  = {\rm \bf I} - \frac{1}{2}(\phi^{\frac{1}{2}} {\rm \bf P}\phi^{-\frac{1}{2}} + \phi^{-\frac{1}{2}}{\rm \bf P} \phi^{\frac{1}{2}}),
\end{equation}
where {\bf I} is the identity matrix, {\bf P} = {\bf D}$^{-1}${\bf A} is the matrix of transition probabilities, {\bf A} is the adjacency matrix, 
{\bf D} is the diagonal matrix of out-degree
(i.e., number of connections of a node directed outward from it) and 
$\phi$ is the Perron-Frobenius eigenvector of {\bf P}. For a strongly connected network defined by {\bf A} (and provided it is aperiodic),
the distribution of random walkers on the network will converge to the stationary distribution given by $\phi$ (see F. Chung,
Ann. Comb. {\bf 9}, 1 (2005) \doi{10.1007/s00026-005-0237-z}).\\

\begin{figure}[ht]
\begin{center}
\includegraphics[width=0.9\linewidth]{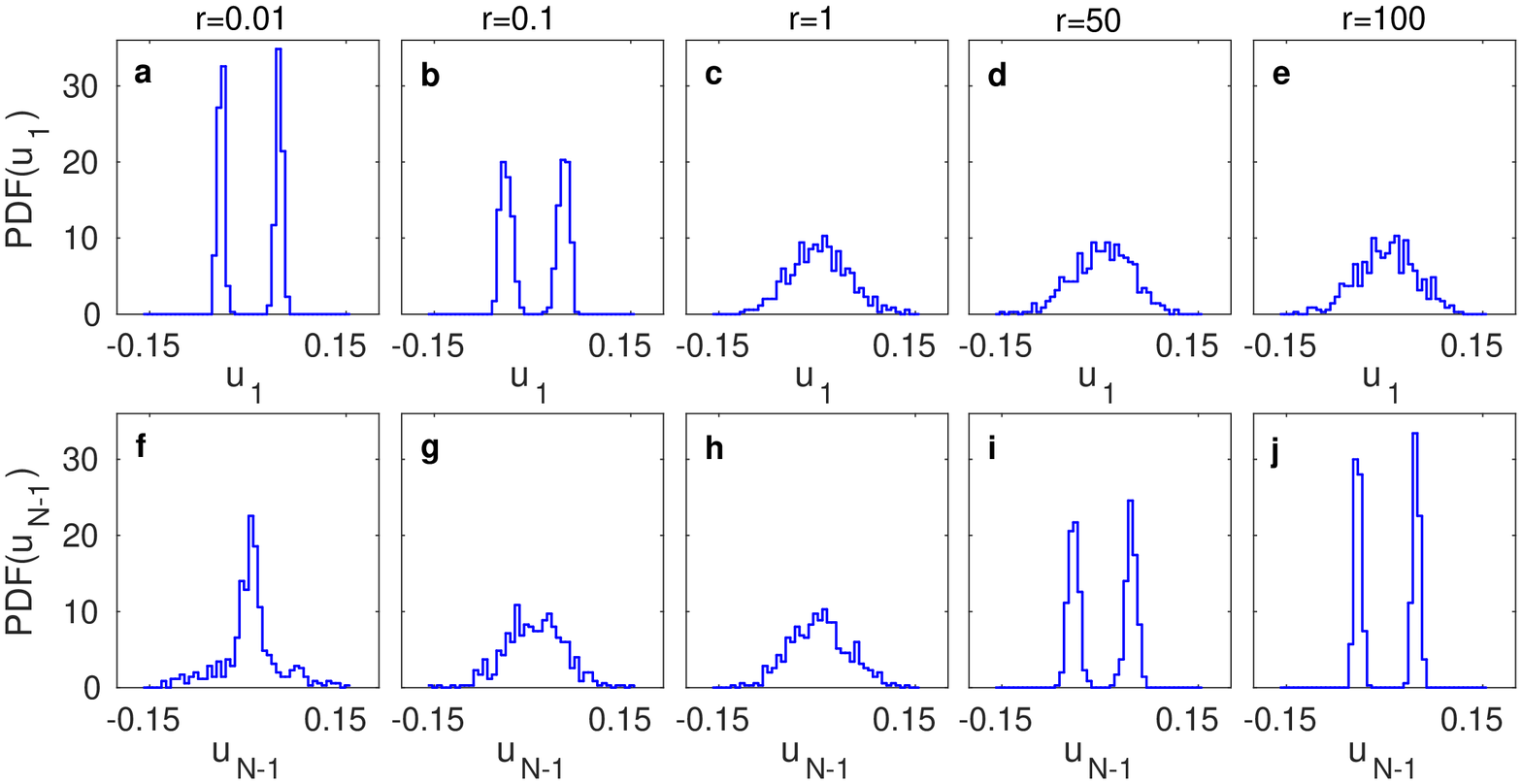}
\end{center}
\caption{Probability distribution for the eigenvector components corresponding to the smallest finite and the largest eigenvalues ( $u_1$ and $u_{N-1}$ respectively) for modular ($r<1$), random ($r=1$) and anti-modular ($r>1$) networks. For modular networks, the components of $u_1$ exhibit a bimodal distribution, whereas, for anti-modular networks the components of $u_{N-1}$ are distributed with a bimodal nature. For all networks shown here
$N=500$, $\langle k \rangle = 20$ and $\sigma_s=0$. }
\label{figs4}
\end{figure}

The distributions of the leading eigenvector $u_{N-1}$ and the eigenector $u_1$ corresponding to the smallest finite
eigenvalue of the normalized symmetric
Laplacian $\mathscr{L}$ for the model networks are shown in Fig.~\ref{figs4}, as the mesoscopic nature of the
network is varied by increasing $r$.  In the main text, these distributions have been suggested as providing signature
for (anti-)modular organization in a network. As can be seen, when $r \ll 1$, such that the networks are modular
in nature, the leading eigenvector has a unimodal distribution while $u_1$ exhibits a bimodal distribution. The reverse
is observed for $r \gg 1$, i.e., when the networks are anti-modular.

\end{document}